\documentclass{appolb_mod}
\usepackage{epsfig}

\def\lsim{\mathrel{\raise.3ex\hbox{$<$\kern-.75em\lower1ex\hbox{$\sim$}}}}
\def\gsim{\mathrel{\raise.3ex\hbox{$>$\kern-.75em\lower1ex\hbox{$\sim$}}}}
\newcommand{\bmk}{\mathbf k}
\newcommand{\bmp}{\mathbf p}
\newcommand{\bmq}{\mathbf q}

\newcommand{\bmA}{\mathbf A}

\newcommand{\bmS}{\mathbf S}

\def\Dsl{\hbox{/\kern-.6000em D}} 

\def\dsl{\,\raise.15ex\hbox{/}\mkern-13.5mu D}
\def\bsigma{\mbox{\boldmath $\sigma$}}

\def\psip#1{\psi_{\mathbf{#1}}}
\def\chip#1{\chi_{\mathbf{#1}}}
\def\bsigma{\mbox{\boldmath $\sigma$}}

\def\abs#1{\left| #1 \right|}
\def\ltap{\ \raise.3ex\hbox{$<$\kern-.75em\lower1ex\hbox{$\sim$}}\ }
\def\gtap{\ \raise.3ex\hbox{$>$\kern-.75em\lower1ex\hbox{$\sim$}}\ }
\def\OMIT#1{}

\def\lsim{\mathrel{\raise.3ex\hbox{$<$\kern-.75em\lower1ex\hbox{$\sim$}}}}
\def\gsim{\mathrel{\raise.3ex\hbox{$>$\kern-.75em\lower1ex\hbox{$\sim$}}}}

\def\O#1#2{\mbox{\boldmath $O$}_{\mbox{\scriptsize\boldmath $#1$},#2}}

\def\Od#1#2{\mbox{\boldmath $O$}^\dagger_{\mbox{\scriptsize\boldmath $#1$},#2}}

\def\msb{{\overline{\rm MS}}}

\newcommand{\nn}{\nonumber}

\begin{document}
\title{Top Pair Production at Threshold and Effective Theories
\thanks{Invited plenary talk at Cracow Epiphany Conference on Heavy Flavors,
January 3-6, 2003.}
}
\author{Andr\'e H.~Hoang
\address{Max-Planck-Institut f\"ur Physik\\
(Werner-Heisenberg-Institut), \\
F\"ohringer Ring 6, \\
80805 M\"unchen, Germany}
}
\maketitle
\begin{abstract}
I give an introduction to the effective field theory description of
top pair production at threshold in $e^+e^-$ annihilation. The impact
of the summation of logarithms of the top quark velocity including
most recent results at next-to-next-to-leading logarithmic order is
discussed. 
\end{abstract}
  
\section{Introduction}

The so-called ``threshold scan'' of the total cross section lineshape of top
pair production constitutes a major part of the top quark physics program at a    
future $e^+e^-$ collider.\,\cite{Teslanlc} In the Standard Model the top
quark width 
$\Gamma_t\approx 1.5$~GeV is much larger than the typical hadronization energy
$\Lambda_{\rm QCD}$. In contrast to the $J/\psi$ or the $\Upsilon$ region, it is
therefore expected that non-perturbative effects are strongly suppressed, and
that the threshold lineshape is a smooth function of the
c.\,m.\,energy. Throughout this talk I will therefore neglect non-perturbative  
effects associated with the hadronization scale $\Lambda_{\rm QCD}$. From
the location of the rise of the cross section a precise measurement of the top
quark mass will be possible, while from the shape and the normalization of the
cross section one can extract the top quark Yukawa coupling $y_t$ (for
a light Higgs), the top width or the strong coupling.
In the past, numerous studies have been carried out to assess the feasibility and
precision for extracting various top quark properties from a threshold run
(see e.g. Ref.\,\cite{TTbarsim} for a recent study).
 
In the threshold region, $\sqrt{s}\simeq 2m_t\pm 10$\,GeV, the top quarks
move with nonrelativistic velocity in the c.m. frame. Let us define $m_t
v^2\equiv\sqrt{s}-2m_t$. We see that parametrically $|v|\lsim
\alpha_s$. Because in the 
loop expansion one encounters terms proportional to $(\alpha_s/v)^n$
($n=0,1,\ldots$) in the amplitude, it is necessary to count $\alpha_s/v$ of
order one and to carry out an expansion in $\alpha_s$ as well as in
$v$. Schematically one 
needs an expansion of $\sigma_{t\bar t}=\sigma(e^+e^-\to t\bar t)$ of the form
\begin{eqnarray}
 R & = & \frac{\sigma_{t\bar t}}{\sigma_{\mu^+\mu^-}}
 \, = \,
 v\,\sum\limits_k \left(\frac{\alpha_s}{v}\right)
\times
 \bigg\{\,1\,\,\mbox{(LO)}\,;
\nonumber \\[2mm] & & \hspace{2cm}
\, \alpha_s, v\,\,\mbox{(NLO)}\,;\, 
 \alpha_s^2, \alpha_s v, v^2\,\mbox{(NNLO)}\,\bigg\}
 \,.
 \label{RNNLOorders}
\end{eqnarray}
The indicated terms are of leading
order (LO), next-to-leading order (NLO), and next-to-next-to-leading order (NNLO).
I call the expansion scheme in Eq.\,(\ref{RNNLOorders}) {\it fixed-order
  perturbation theory\,}, although it involves summations of the terms 
 proportional to $(\alpha_s/v)^n$. It can be implemented systematically using the
factorization properties of non-relativistic QCD
(NRQCD)~\cite{CaswellLepage}. The NNLO QCD corrections 
to the total cross section were calculated some time ago in
Ref.\,\cite{Hoang1,Melnikov1,Yakovlev1,Beneke1,Nagano1,Hoang2,Penin1}.
Surprisingly, the corrections were found to be as large as the next-to-leading
order (NLO) QCD corrections, and from the residual scale dependence in the NNLO
result, the normalization of the cross section was estimated to have at least
$20\%$ theoretical uncertainty.\,\cite{Hoang3} It was concluded that the top
quark mass  in a threshold mass scheme can be determined with a precision of
$200$~MeV or 
better from the shape of the cross section.\,\cite{Hoang3} However, the
large NNLO QCD corrections to the normalization of the cross section
jeopardized competitive measurements of the top width, strong top coupling, or
the top Yukawa coupling. Moreover, the large NNLO corrections 
seemed to indicate that, despite the perturbative nature of the $t\bar t$
system, high precision computations might not be
feasible. The determination of higher orders would be
necessary to clarify the feasibility of the fixed-order
approach. Figure\,\ref{figtopplots}a  
shows the vector-current-induced cross section $\sigma(e^+e^-\to\gamma^*\to
t\bar t)$ at LO, NLO and NNLO in fixed-order perturbation theory
for typical choices of parameters and 
renormalization scales. (Top threshold production mediated by the axial vector
current is suppressed by $v^2$ and small and not relevant here for the
discussions on the convergence of the perturbative expansion. See
e.g. Refs.\,\cite{Hoang3,Hoang4}.)   

One way to understand the large scale uncertainties and the large size of the
fixed-order 
NNLO QCD corrections is to recall that the perturbative $t\bar t$ threshold
dynamics is governed by vastly different energy scales, the top mass ($m_t\sim
175$\,GeV), the top three-momentum ($\bmp\simeq m v\simeq 25$\,GeV) and the top
kinetic or potential energy ($E\simeq m v^2\simeq 4$\,GeV). This hierarchy of
scales is the basis of NRQCD factorization\,\cite{CaswellLepage},
which separates hard 
and non-relativistic effects, and which can be implemented in the fixed-order 
scheme shown above. 
However, in fixed-order perturbation theory NRQCD matrix elements involve
logarithmic terms such as 
\begin{equation}
\ln\Big(\frac{\mu^2}{m_t^2}\Big)\,,\qquad
\ln\Big(\frac{\mu^2}{\bmp^2}\Big)\,,\qquad
\ln\Big(\frac{\mu^2}{E^2}\Big)\,,
\label{logarithms}
\end{equation}
which cannot be rendered small for a single choice of the renormalization
scale $\mu$. For example, since $E\sim 4$\,GeV, one finds
$\alpha_s(m_t)\ln(m_t^2/E^2)\simeq 0.8$ for $\mu=m_t$ which is of order unity,
and fixed-order perturbation theory becomes unreliable. Moreover, at higher
orders, fixed-order perturbation theory cannot distinguish the scale at
which to evaluate $\alpha_s$.  Mistaking an $\alpha_s(m_t)$ for an
$\alpha_s(m_t v^2)$ is a difference of a factor of two. Both problems cannot
be addressed systematically within the framework of fixed-order perturbation
theory. But they can be 
avoided in a framework that allows for renormalization group improved
perturbative computations, where {\it all\,} logarithmic terms are summed
through renormalization group equations. Thus a better expansion should have
the schematic form 
\begin{eqnarray}
 R & = & \frac{\sigma_{t\bar t}}{\sigma_{\mu^+\mu^-}}
 \, = \, v\,\sum\limits_k \left(\frac{\alpha_s}{v}\right)
 \sum\limits_i \left(\alpha_s\ln v \right)^i 
\nonumber\\ & &
\times
 \bigg\{\,1\,\,\mbox{(LL)};\, \alpha_s, v\,\mbox{(NLL)};\, 
 \alpha_s^2, \alpha_s v, v^2\,\mbox{(NNLL)}\,\bigg\}
 \,,
 \label{RNNLLorders}
\end{eqnarray}
where the indicated terms are of leading logarithmic (LL), next-to-leading
logarithmic (NLL), and next-to-next-to-leading logarithmic (NNLL) order. 
I have used the expression ``better expansion'' because in QCD computations
one should favor expansion schemes where logarithmic terms are summed into
coefficients and not contained in isolated form in matrix elements. This is
even true in cases 
where the non-logarithmic corrections are sizeable too, because possible 
cancellations between large logarithmic and large non-logarithmic terms can be
unphysical. To accomplish a renormalization group improved calculation one
needs to employ a more sophisticated effective field theory approach than the 
one that is used for the fixed-order computations.  

In this talk I give a basic introduction to ``velocity NRQCD''
(vNRQCD)~Ref.\,\cite{LMR,amis,amis2,Hoang4} (for a review see also
Ref.\,\cite{Reviewhoang}), which is an effective theory for heavy
non-relativistic quark pairs that, through renormalization, sums
all logarithms involving ratios of the scales $m_t$, $|\bmp|$
and $E$.  It is assumed that $E\gg\Lambda_{\rm QCD}$. The
program is similar in spirit to let's say summing QCD logarithms 
of $M_W/m_b$ for the electroweak Hamiltonian describing b-quark
decays. However, in the non-relativistic case the program is more complicated
because the 
structure of the relevant degrees of freedom in the effective theory action is
more involved and because the two low energy scales $\bmp\sim mv$ and $E\sim
mv^2$ are correlated through the heavy quark equation of motion $E=\bmp^2/m$.
At first sight it seems impossible to keep this correlation of scales and
to render the logarithms in the effective theory matrix
elements shown in Eq.\,(\ref{logarithms}) small (which is equivalent to saying
that all logarithms are summed into Wilson coefficients of the effective
theory). In the effective theory this problem is dealt with by using two
renormalization group scales in the effective Lagrangian: $\mu_S$ for soft
($\sim mv$) fluctuations and $\mu_U$ for ultrasoft ($\sim mv^2$) fluctuations.
The matrix elements of the
effective theory then only contain the logarithms $\ln(\mu_S^2/\bmp^2)$ and  
$\ln(\mu_U^2/E^2)$. Both renormalization scales are related by
$\mu_U=\mu_S^2/m_t$. It is therefore convenient to define $m\nu\equiv\mu_S$
and $m\nu^2\equiv\mu_U$, where $\nu$ is a dimensionless parameter. All
renormalization group equations of the effective theory are expressed in terms
of $\nu$. Running from $\nu=1$ to $\nu\sim v$, where $v$ is the typical top
quark velocity, sums all logarithms of $v$ and minimizes both
$\ln(\mu_S^2/\bmp^2)$ and $\ln(\mu_U^2/E^2)$ in the matrix
elements. (Logarithms involving $m_t$ are minimized by matching QCD onto the
effective theory at $\nu=1$.) The correlation of both renormalization scales
is essential for the correct summation of logarithms.\,\cite{ManoharSoto}.

In dimensional regularization the factors
of $\mu_S^\epsilon$ and $\mu_U^\epsilon$ multiplying each operator in the
effective 
theory action are uniquely determined from its mass dimension and $v$ power
counting. In this way the scheme indicated in Eq.\,(\ref{RNNLLorders}) can
indeed be achieved.
 
With this setup and after having formulated the effective theory action the
program is similar the one for summing QCD logarithms for the electroweak
Hamiltonian:
\begin{enumerate}

\item Matching computation of the coefficients $C_i(\nu)$ of the effective
  theory operators (including external currents) at $\nu=1$ in a
perturbation series in $\alpha_s(m_t)$.

\item Computation of the anomalous dimensions of the operators of the
  effective action and scaling of $C_i(\nu)$ from $\nu=1$ to
  $\nu=v_0\simeq v\simeq C_F\alpha_s$ using the renormalization group.

\item Computation of the cross section using the effective Lagrangian and
  the currents renormalized at the low scale $\nu=v_0$.

\end{enumerate}

In this talk I show how the vNRQCD effective Lagrangian is
constructed, and I also spend some time 
on the points 1.-3. described above including new recent results for the total
top pair production cross section at threshold in $e^+e^-$ annihilation. At
the end I will briefly discuss the status of top pair production at
threshold. 

\section{The vNRQCD Lagrangian}
\label{sectionvNRQCD}

The physical system we wish to describe is that of a heavy quark and
antiquark with mass $m$, and energies $E\sim mv^2$, and momenta ${\bf p}\sim
mv$ in the c.m.\,system where $v\ll 1$. The degrees of freedom from
which we have to build the effective action can be identified from the
relevant momentum regions that can be found in an asymptotic expansion of
non-relativistic scattering diagrams in the c.m.\,frame.\,\cite{Beneke} The
regions include 
hard modes with momenta $(k_0,\bmk)\sim (m,m)$, soft modes with momenta
$\sim (m v,m v)$, potential modes with momenta $\sim (m v^2,m v)$ and
ultrasoft modes with momenta $\sim (m v^2,m v^2)$.
Fluctuations with off-shell momenta are integrated out because they do not
resonate, while for modes that can resonate fields are introduced from which
we build the effective Lagrangian. Resonating heavy quarks can only live in
the potential regime, while massless modes can live either in the soft
or the ultrasoft regime. Thus, the effective Lagrangian is build from 
heavy potential quarks and antiquarks ($\psi_\bmp$, $\chi_\bmp$), soft gluons,
ghosts, and massless quarks ($A_q^\mu$, $c_q$, $\varphi_q$) and ultrasoft
gluons, ghosts, and massless quarks ($A^\mu$, $c$, $\varphi_{us}$).
The ultrasoft gluons are the gauge partners of momenta $\sim mv^2$, while soft 
gluons are the gauge partners of momenta $\sim mv$. This classification is, of
course, only meaningful in the c.m.\,frame for a non-relativistic heavy
quark-antiquark pair. Double counting is avoided since ultrasoft gluons
reproduce only the physical gluon poles where $k^0\sim 
{\bmk}\sim mv^2$, while soft gluons only have poles with $k^0\sim {\bmk}\sim 
mv$. It is essential that both soft and ultrasoft gluons are
included at all scales below $m$ because the heavy quark equation of motion
correlates the soft and ultrasoft scales. We will also see later in
Sec.\,\ref{sectionanomdim}, when I show new results for the 3-loop (NNLL) anomalous
dimension for the $t\bar t$ production current, that both soft and ultrasoft
running can in general feed into anomalous dimensions induced by potential
loops. The dependences on soft energies and momenta of potential and soft
fields appear as labels on the fields, while only the lowest-energy ultrasoft
fluctuations are associated by an explicit coordinate dependence. Formally
this is achieved by a phase redefinition
for the potential and soft fields~\cite{LMR}
\begin{eqnarray}
  \phi(x) \to \sum_k e^{-ik\cdot x} \phi_k(x) \,,
\end{eqnarray}
where $k$ denotes momenta $\sim mv$ and $\partial^\mu \phi_k(x)\sim mv^2
\phi_k(x)$. This means that the effective Lagrangian contains sums over
fields with soft indices, which also build up potential and soft loop
integrations 
when we renormalize the theory or compute matrix elements.

The effective vNRQCD Lagrangian for a $t\bar t$ angular momentum S-wave and
color singlet state has terms\,\cite{LMR,amis,amis2}  
\begin{eqnarray}
\lefteqn{
{\cal L}  = 
 \sum_{\mathbf p}\,\bigg\{
   \psi_{\bmp}^\dagger   \bigg[ i D^0 \!-\! {\left({\bf p}\!-\!i{\bf D}\right)^2
   \over 2 m_t} +\frac{{\mathbf p}^4}{8m_t^3} + \ldots \bigg] \psi_{\bmp}
 + (\psi \to \chi)\bigg\} -{1\over 4}G^{\mu\nu}G_{\mu \nu}
}
\nonumber \\[2mm] && 
 - \mu_S^{2\epsilon}g_s^2 \sum_{{\bmp},{\bmp^\prime},q,q^\prime,\sigma} \bigg\{ 
  \frac{1}{2}\, \psi_{\bmp^\prime}^\dagger
 [A^\mu_{q^\prime},A^\nu_{q}] U_{\mu\nu}^{(\sigma)} \psi_{\bmp}
 + (\psi \to \chi) + \ldots\, \bigg\}
\nonumber \\[3mm] &&  
 + \sum_{p} \abs{p^\mu A^\nu_p -
 p^\nu A^\mu_p}^2 + \ldots
-\,\sum_{{\bmp},{\bmp^\prime}}
 \mu_S^{2\epsilon} V({\bmp,\bmp^\prime})\,\psi_{\bmp^\prime}^\dagger \psi_{\bmp}
   \chi_{-\bmp^\prime}^\dagger \chi_{-\bmp}
\nonumber \\[2mm] &&  
+\sum_{{\bmp},{\bmp^\prime}} \frac{2i\mu_S^{2\epsilon}{\cal V}_c}
  {(\bmp^\prime-\bmp)^4} f^{ABC}
  {(\bmp-\bmp^\prime)}.(\mu_U^\epsilon g_u {\bmA}^C) [\psi_{\bmp^\prime}^\dagger
  T^A \psi_{\bmp} \chi_{-\bmp^\prime}^\dagger \bar T^B \chi_{-\bmp} ] 
+\ldots
\,,
\label{vNRQCDLagrangian}
\end{eqnarray} 
where color and spin indices have been suppressed and $g_s\equiv g_s(m_t\nu)$,
$g_u\equiv g_s(m_t\nu^2)$. All coefficients are functions of the renormalization
parameter $\nu$, and all explicit soft momentum labels are summed. The
covariant 
derivative in the first line contains only the ultrasoft gluon field. There are
4-quark potential-like interactions of the form ($\bmk=(\bmp-\bmp^\prime)$) 
\begin{eqnarray}
 V({\bmp},{\bmp^\prime}) & = & 
 \frac{{\cal V}_c}{\bmk^2}
 + \frac{{\cal V}_r({\bmp^2 + \bmp^{\prime 2}})}{2 m_t^2 \bmk^2}
 + \frac{{\cal V}_2}{m_t^2}
 + \frac{{\cal V}_s}{m_t^2}{\bmS^2}\,,
\label{vNRQCDpotential}
\end{eqnarray}
where $\bmS$ is the total $t\bar t$ spin operator. Note that the momentum
structure of the operators satisfies the on-shell conditions and hermiticity.
There is still some freedom in
the choice of the operator basis for the potential interactions shown in
Eq.\,(\ref{vNRQCDpotential}), which can affect the matching conditions and the 
anomalous dimensions. In Refs.\,\cite{Hoang4,hmst,Hoang5} we also used a
potential interaction term 
of the form $1/(m|{\bmk}|)$, while in Ref.\,\cite{Hoang6} the order 
$1/(m|{\bmk}|)$ potentials were implemented only by  potential-like
interactions with additional sums over intermediate indices. 

At NNLL order for the total
cross section the coefficient ${\cal V}_c$ of the $1/\bmk^2$  potential has to be
matched at two loops~\cite{hms1} because it contributes at the LL level, whereas
the coefficients of the order $1/m_t^2$ potentials have to matched at the Born
level~\cite{amis}. The $1/(m|{\bmk}|)$-type potentials are of order
$\alpha_s^2$ and have to be matched at two loops.\,\cite{Hoang5} 
There are also 4-quark interactions  with the radiation of an
ultrasoft gluon  (last line) and interactions between quarks and soft gluons
(second line). Due to 
momentum conservation at least two soft gluons are required. The potential
terms shown in Eq.\,(\ref{vNRQCDpotential}), and time-ordered
products of soft interactions contribute to the potentials that describe the
instantaneous interactions between the top quarks. 
For example, the time-ordered product of two soft interactions at leading order
leads to the potential interaction~\cite{hms1}
\begin{eqnarray}\lefteqn{
 \tilde V_{\rm soft}(\bmp,\bmq) 
 \, = \,
 -\,\frac{4\pi C_F\, \alpha_s(\mu_S)}{\bmk^2}\, \Bigg\{\,
 \frac{\alpha_s(\mu_S)}{4\pi}\,\bigg[\,
 -\beta_0\,\ln\Big(\frac{\bmk^2}{\mu_S^2}\Big) + a_1
 \,\bigg]} \nonumber\\[2mm] 
& & \quad + \bigg(\frac{\alpha_s(\mu_s)}{4\pi}\bigg)^2\,\bigg[\,
 \beta_0^2\,\ln^2\Big(\frac{\bmk^2}{\mu_S^2}\Big)  
 - \Big(2\,\beta_0\,a_1 +
 \beta_1\Big)\,\ln\Big(\frac{\bmk^2}{\mu_S^2}\Big) + a_2 \,\bigg]
 \,\Bigg\} \,,\quad
 \label{VCoulombspoft}
\end{eqnarray}
where the $\beta_i$ are the coefficients of the QCD beta-function and the $a_i$
were determined some time ago in Ref.\,\cite{Schroder}. 
The sum of Eq.\,(\ref{VCoulombspoft}) and the first term in
Eq.\,(\ref{vNRQCDpotential}) agrees with the static 
potential\,\cite{PSstatic}\footnote{The
  statement made in Refs.\,\cite{hms1,Reviewhoang} that the vNRQCD $1/\bmk^2$
  potential at 
  NNLL order differs from the static potential has shown to be incorrect in 
  Ref.\,\cite{Hoang5}.}  
and constitutes the complete $1/\bmk^2$ potential at NNLL order. 
The Lagrangian also contains more complicated interactions describing
4-quark potential-like interactions and 4-quark interactions with soft or
ultrasoft gluons that contain additional sums over intermediate indices.
Their form is partly fixed by the renormalizability of the
theory~\cite{Hoang5}, but there is still some freedom in the choice of their
form. In this talk I use 
the conventions of Ref.\,\cite{Hoang6} were all $1/(m|{\bmk}|)$-type
potentials are represented by
such 4-quark sum operators. I do not display their expressions here and refer to
Ref.\,\cite{Hoang6} where they are collected in an appendix. 

One might wonder why
the effective theory contains both soft and ultrasoft degrees of freedom at
the same time, since, naively, one might argue that for $v\ll 1$ the soft
fields fluctuate at a much smaller length scale than the ultrasoft fields.  
That both soft and ultrasoft degrees of freedom have to be present at the same
time is at the heart of the effective theory construction and is related to the
fact that the dispersion relation of the heavy quarks, $E=\bmp^2/m$ introduces
a correlation between the soft and ultrasoft scales. The presence of soft and
ultrasoft degrees of freedom also allows for the simultaneous summation of
logarithms of $\bmp$ and $E$, which I mentioned in the introduction, after the
theory is renormalized. I will come back to this point in the next section. 

An important point to mention is that the factors of $\mu_U^\epsilon$ and
$\mu_S^\epsilon$ multiplying the operators in the effective Lagrangian are
uniquely determined by the mass dimension and the $v$ power counting in
$d=4-2\epsilon$ dimensions~\cite{amis2}. Each field in the action is assigned
a certain scaling with $v$ to keep its kinetic term of order $v^0$. This
results for example in $\psi_{\bf p}\sim (mv)^{3/2-\epsilon}$, 
$A^\mu\sim (mv^2)^{1-\epsilon}$, and $A_q^\mu\sim (mv)^{1-\epsilon}$. Then,
since $D^\mu\sim mv^2$, the renormalized combination $g_u A^\mu$ must be
multiplied by $\mu_U^\epsilon \sim (m\nu^2)^{\epsilon}$ for $\nu\sim v$ so that
this gluon term also scales consistently as $mv^2$. Applying the same
procedure to all interactions results in the factors of $\mu_U^\epsilon$ and
$\mu_S^\epsilon$ shown for example in Eq.\,(\ref{vNRQCDLagrangian}). Note that
this procedure is an integral part of the theory and automatically leads to
the correlation of the two renormalization scales,
$\mu_U\propto\mu_S^2/m$.

To incorporate the effect of the large top quark width we include in the 
effective Lagrangian the operators
\begin{eqnarray} \label{Gtop}
 \delta {\cal L} = \sum_{\bmp} \psip{p}^\dagger \: \frac{i}{2} \Gamma_t\: 
  \psip{p} \ +\  
  \sum_{\bmp}   \chip{p}^\dagger\: \frac{i}{2} \Gamma_t\: \chip{p} \,,
\end{eqnarray}
where $\Gamma_t$ is the total on-shell top quark width. The typical
energy of the top quarks at threshold is $E\sim mv^2\sim 
4\,{\rm GeV}$, and in the Standard Model one has
$\Gamma_t=1.43$\,GeV\,$\sim E$. Thus, the propagator for a single 
top (or antitop) with momentum $(p^0,{\bmp})$ is
\begin{eqnarray}
  \frac{i}{p^0 - {\bmp}^2/(2m) + i\Gamma_t/2 + i\epsilon} \,,
\end{eqnarray}
which gives a consistent NLO treatment of electroweak
effects.\,\cite{nonfactorizable}   
The complete NNLL order treatment of electroweak effects for the total cross
section is currently unknown. Although it is not expected that the unknown
electroweak corrections are beyond the few percent level~\cite{Hoang2}, this
has to be kept in mind when I discuss the NNLL QCD corrections.

Besides the interactions contained in the effective Lagrangian that describe
the dynamics of the $t\bar t$ pair we also need external currents that
describe the production of the top quarks. For $e^+e^-$ annihilation 
these currents are induced by the exchange of a virtual photon or a Z boson.
At NNLL order we need the vector S-wave currents  ${\bf
J}^v_{\bmp}= c_1(\nu) \O{p}{1} + c_2(\nu) \O{p}{2}$, where
\begin{eqnarray}\label{Ov}
 \O{p}{1} & = & {\psip{p}}^\dagger\, \bsigma(i\sigma_2)\, {\chip{-p}^*} \,, 
   \\[2mm]
 \O{p}{2} & = & \frac{1}{m^2}\, {\psip{p}}^\dagger\, 
    \bmp^2\bsigma (i\sigma_2)\, {\chip{-p}^*} \,, \nn
\end{eqnarray} 
and the axial-vector P-wave current  ${\bf J}^a_{\bmp}=
c_3(\nu) \O{p}{3} $, where
\begin{eqnarray}\label{Oa}
 \O{p}{3} & = & \frac{-i}{2m}\, {\psip{p}}^\dagger\, 
      [\,\bsigma,\bsigma\cdot\bmp\,]\,(i\sigma_2)\,
   {\chip{-p}^*} \,. 
\end{eqnarray}
Note that the currents have a soft momentum index. The currents $\O{p}{2}$ and 
$\O{p}{3}$ lead to contributions in the total cross section that are
$v^2$-suppressed with respect to those of the current $\O{p}{1}$.
Thus, at NNLL order, two-loop matching is needed for $c_1$ and Born level
matching for $c_2$ and $c_3$. Note that the two-loop matching result for
$c_1(\nu=1)$ depends on the choice of the operator basis used in the effective
Lagrangian.  

\section{Anomalous Dimensions and Renormalization Group Scaling}
\label{sectionanomdim}

To sum the logarithmic terms $(\alpha_s\ln\nu)^i$ at NNLL order, as indicated
schematically in Eq.\,(\ref{RNNLLorders}), one needs to determine the anomalous
dimensions of the potentials and currents that contribute to the cross section
at the appropriate order. The computation of the anomalous dimensions in the
effective theory works just as for the effective weak Hamiltonian.
The coefficient of the $1/\bmk^2$ (Coulomb)
potential, ${\cal V}_c$, needs to be determined at the three-loop level since
it contributes already at leading order. The results are available in
Ref.\,\cite{Hoang5}. The anomalous dimensions of the $1/m_t^2$ potentials, 
${\cal V}_{r,2,s}$ only need to be determined at one-loop because the
$1/m_t^2$ potentials are suppressed by $v^2$ and they are only needed at LL
order. The results have been given in Refs.\,\cite{amis,Hoang5}. 
(See also Refs.\,\cite{PSstatic,PS1}.) 
For the $1/(m|\bmk|)$ potentials, on the other hand, the anomalous dimensions have
to determined at two-loops. Also this work has been achieved.~\cite{Hoang5}
For the current $\O{p}{1}$ one needs to compute the anomalous dimension at
three loops and for $\O{p}{2}$ and $\O{p}{3}$, they are needed at the one-loop
level. The LL anomalous dimensions of $\O{p}{2}$ and $\O{p}{3}$ were computed
in Refs.\,\cite{Hoang4,hmst}. I will not give any details on these results 
in this talk since they are already available in the literature for some
time. 

Here I want to discuss some aspects of the anomalous dimension of the current
$\O{p}{1}$, which is currently only fully known to NLL
order\,\cite{LMR,Pinedacurrent,Hoang5}, but for which I
have recently obtained a partial NNLL order result.\,\cite{Hoang6} The NNLL
order results I will discuss are also of conceptual interest as far as the
construction and the renormalizability of the effective theory is concerned. 

To obtain the anomalous dimension of $\O{p}{1}$, one needs to determine the
renormalization constant of the current. Using dimensional regularization and
the $\msb$ scheme one can define the unrenormalized Wilson
coefficient as \begin{equation}
c_1^0 \, = \, c_1 \, + \, \delta c_1
 \, = \, Z_{c_1}\,c_1
\,,
\end{equation}
and one can write the renormalization constant as
\begin{equation}
Z_{c_1} \, = \, 1 \, + \,
\frac{\delta z_{c_1}^{\rm NLL}}{\epsilon}\, + \,
\Big(\,
\frac{\delta z_{c_1}^{\rm NNLL,2}}{\epsilon^2} \, + \,
\frac{\delta z_{c_1}^{\rm NNLL,1}}{\epsilon}\,\Big)\, + \,
\ldots
\,.
\end{equation}
At LL order there are no UV divergences that have to be absorbed into the
current. This means that at LL order the Wilson coefficient $c_1$ does
not run. 

To determine the N$^k$LL order renormalization constant of the 
current $\O{p}{1}$ one has to compute the overall UV divergences
of quark-antiquark-to-vacuum {\it on-shell} matrix elements of spin-triplet
S-wave currents at ${\cal O}(\alpha_s^{k+1})$. All lower order
UV subdivergences have to be subtracted by lower order counterterms. It is
mandatory to use matrix elements where the external quarks are on-shell
because for off-shell quarks one can obtain UV divergences which do not belong
to the current. This is because we have defined our operator basis
in the on-shell limit, particularly for the potentials. Most of the
time one does not have to worry about this issue, but such off-shell
UV divergences in fact exist at NNLL order. 

Let me first review the NLL result which was first obtained in
Ref.\,\cite{LMR}. 
%
%
\begin{figure}[t] 
\begin{center}
 \leavevmode
 \epsfxsize=4cm
 \leavevmode
 \epsffile[165 160 485 575]{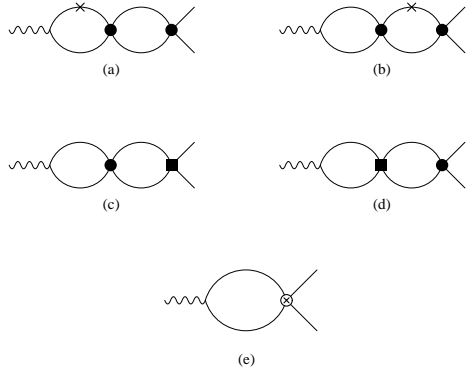}
 \vskip  0.0cm
 \caption{
Order $\alpha_s^2$ vertex diagrams diagrams for the computation of the NLL
anomalous dimension of $c_1$. The dot and box represent single insertions of
the LL $1/\bmk^2$ and the $1/m_t^2$ suppressed potentials, respectively. The
circled cross denotes a single insertion of the $1/(m|\bmk|)$-type potentials.
Insertions of these potential involve only a one-loop diagram because
the coefficients of the $1/(m|\bmk|)$-type potentials are of order
$\alpha_s^2$.  
 \label{fignllvertex} }
\end{center}
\end{figure}
The relevant order $\alpha_s^2$ vertex diagrams are shown in
Fig.\,\ref{fignllvertex}. There are no one-loop subdivergences that have to be
subtracted. The computation is complicated technically by the fact that 
there are IR-divergent Coulomb phases for
on-shell external quarks which have to be distinguished from the
UV divergences that go into the renormalization of the current. At NLL order
this distinction can be carried out easily because only $1/\epsilon$
singularities occur. From the UV divergences one obtains
\begin{eqnarray}
\lefteqn{
\delta z_{c_1}^{\rm NLL} \, = \,
 -\:\frac{{\cal V}_c^{(s)}(\nu)
  }{ 64\pi^2} \bigg[ \frac{ {\cal V}_c^{(s)}(\nu) }{4 }
  +{\cal V}_2^{(s)}(\nu)+{\cal V}_r^{(s)}(\nu)
   + {\bf S}^2\: {\cal V}_s^{(s)}(\nu)  \bigg] }
   \nn\\
  && +\: \alpha_s^2(m\nu)\,\bigg[ \frac{C_F}{2}(C_F-2\,C_A)\bigg] 
     +  \alpha_s^2(m\nu)\,\bigg[  
     3 {\cal V}_{k1}^{(s)}(\nu) + 2 {\cal V}_{k2}^{(s)}(\nu) \bigg] 
\,,
\label{zc1}
\end{eqnarray}
where ${\bf S}^2=2$ is the squared quark total spin operator for the
spin-triplet configuration. The terms in the second line arise from a single
insertion of the $1/(m|\bmk|)$-type potentials.
Using that $c_1^0$ is renormalization group
invariant, one can derive the NLL order anomalous dimension of $c_1$,
\begin{eqnarray}\label{c1anomdim}
 \nu \frac{\partial}{\partial\nu} \ln[c_1(\nu)] & = &
\gamma_{c_1}^{\rm NLL}(\nu) + 
\gamma_{c_1}^{\rm NNLL}(\nu) +\ldots
\,,
\nonumber\\[2mm]
\gamma_{c_1}^{\rm NLL}(\nu) & = & 4\, \delta z_{c_1}^{\rm NLL}
\,.
\end{eqnarray}
To solve the NLL renormalization group equation one needs to know the LL 
solutions for all the couplings that appear on the RHS of
Eq.\,(\ref{c1anomdim}) (i.e. which mix into $c_1$). From the fact that all
higher order $1/\epsilon^n$ ($n=1,2,\ldots$) divergences have to cancel in the 
anomalous dimension (see e.g. Ref.\,\cite{Collins1}), one can also obtain the 
NNLL order $1/\epsilon^2$ coefficient $\delta z_{c_1}^{\rm NNLL,2}$ of the
renormalization constant $Z_{c_1}$.~\cite{Hoang6} 

The determination of the NNLL order anomalous dimension proceeds along the
same lines. One can distinguish between contributions from two classes.
The first class, called ``mixing contributions'', arises from the terms in
Eq.\,(\ref{zc1}) due to the NLL order running of the couplings on the RHS of
Eq.\,(\ref{c1anomdim}). 
The second class involves the computation of three-loop vertex diagrams
with potential loops and either soft or ultrasoft loops that require new $c_1$
counterterms. I call this second class ``non-mixing contributions''. It leads
to genuinely new contributions in the anomalous dimension of $c_1$. By power
counting there are no contributions from diagrams with three potential loops
or which have both soft and ultrasoft loops. I have determined the non-mixing
contributions recently in Ref.\,\cite{Hoang6} and I will discuss the computation
and the results in some detail in the following. 

At NNLL order the determination of the UV divergences from on-shell three-loop
vertex diagrams is complicated because there 
are overlapping IR and UV divergences which are difficult to separate from
each other.  An efficient way to avoid these complication is to consider
4-loop current correlator graphs rather than the vertex diagrams. The
correlator graphs are 
obtained from closing the external quark lines of the vertex diagrams with an
additional insertion of the current. 
%
%
\begin{figure}[t] 
\begin{center}
 \leavevmode
 \epsfxsize=12cm
 \leavevmode
 \epsffile[45 245 545 575]{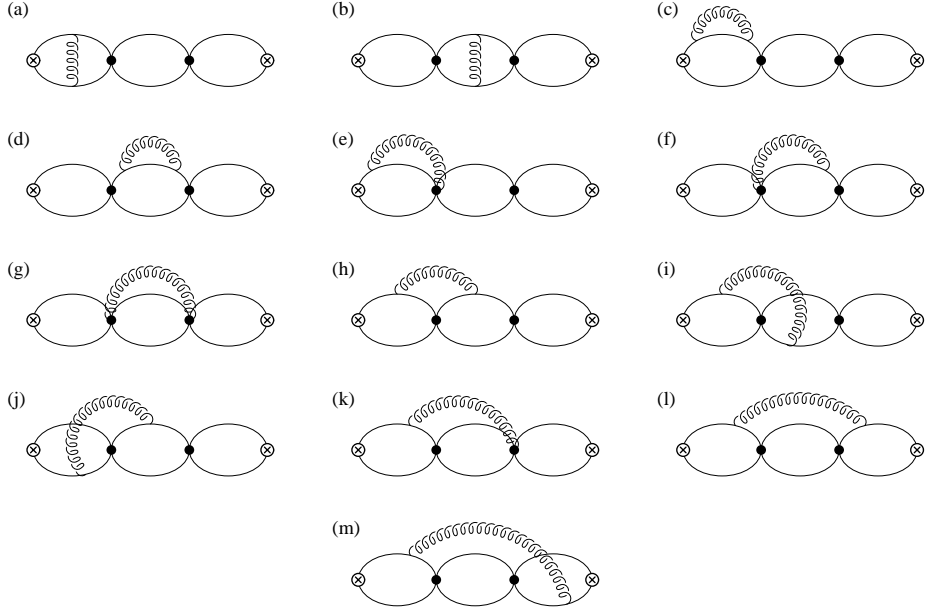}
 \vskip  0.2cm
 \caption{
Four-loop graphs with an ultrasoft gluon for the calculation of the ultrasoft
non-mixing contributions of 
the NNLL anomalous dimension of $c_1$. The ultrasoft gluon is attached to
the top quark or to potential-like 4-quark interactions.
 \label{figusoftcorrelators} }
\end{center}
\end{figure}
In Fig.\,\ref{figusoftcorrelators} I have exemplarily displayed the four-loop
correlator diagrams with an ultrasoft gluon (loop).
The imaginary part of the correlator
graphs is proportional to the squared matrix elements from which all
IR-divergent Coulomb phases drop out automatically.  The three-loop (NNLL) 
renormalization constant of the current is then obtained from
the three-loop subdivergences of the correlator diagrams
that remain after the one- and  two-loop subdivergences have been  
subtracted. For dimensional reasons there are in fact no four-loop overall
divergences in dimension regularization. 

Due to lack of space I cannot give all the details of the computation and the
full result here and refer to Ref.\,\cite{Hoang6}. But I would 
like to discuss some interesting conceptual aspects of the results. First,   
the NNLL order $1/\epsilon^2$ coefficient $\delta z_{c_1}^{\rm NNLL,2}$ I have
obtained by the computations agrees with the NLL prediction based on
renormalization group invariance 
mentioned above. This is a non-trivial check for the consistency of the
effective theory under renormalization, since it requires that the
interactions between the relevant degrees of freedom are encoded correctly
in the structure of the effective Lagrangian. This shows in particular, that
the renormalizability of the effective theory requires that soft and ultrasoft
degrees of freedom have to be present for all scales below the quark mass
$m_t$. In addition, the results for the diagrams in
Fig.\,\ref{figusoftcorrelators}, which involve the soft and the ultrasoft
renormalization scales, $\mu_S$ and 
$\mu_U$, show that the correlation $\mu_U=\mu_S^2/m_t$ I have mentioned
previously is required by the renormalizability of the effective theory. As an
example, let us consider the  
contribution to $\delta Z_{c_1}^{\rm NNLL,2}$ induced by the diagram 
Fig.\,\ref{figusoftcorrelators}g which, when the correlation between $\mu_S$
and $\mu_U$ is neglected, has the form 
\begin{eqnarray} \label{ABresults}
\frac{\alpha_s(m_t\nu)^2\alpha_s(m_t\nu^2)}{\pi}
\frac{C_A^2C_F^2}{4}\,\bigg[\,
 -\,\frac{1}{12\epsilon^2}-
\frac{1}{6\epsilon}\,\bigg[\,
\ln 2+\frac{7}{6}-
\ln\Big(\frac{m \, \mu_U}{\mu_S^2}\Big)
\,\bigg]
\,.
\end{eqnarray}
The dependence on $\ln\mu_S$ and $\ln\mu_U$ in the $1/\epsilon$ piece that
contributes to the NNLL anomalous dimension of $c_1$  vanishes only if  
the correlation $\mu_U=\mu_S^2/m$ is accounted for. This shows that the
correlation of the soft and ultrasoft scales in the effective theory is
required by physical reasons and not just imposed by hand.

Although the computation of the NNLL anomalous dimension of $c_1$ is not
completed yet, it
is instructive to compare the numerical size of the non-mixing
NNLL contributions with the NLL ones in the running of $c_1$. 
Let us write the solution of Eq.\,(\ref{c1anomdim}) for $\nu< 1$ as
\begin{eqnarray}
 \ln\Big[ \frac{c_1(\nu)}{c_1(1)} \Big] & = &
\xi^{\rm NLL}(\nu) + 
\Big(\,
\xi^{\rm NNLL}_{\rm m}(\nu) + \xi^{\rm NNLL}_{\rm nm}(\nu)
\,\Big) + \ldots
\,.
\label{c1solution}
\end{eqnarray}
where $\xi^{\rm NNLL}_{\rm nm}$ ($\xi^{\rm NNLL}_{\rm m}$) refers to the NNLL
non-mixing (mixing) contributions.  
\begin{table}[t]
\begin{center}
\begin{tabular}{|c||c|c||c|c|}
\hline
 & \multicolumn{2}{|c||}{$m=175$\,GeV}
 & \multicolumn{2}{|c|}{$m=4.8$\,GeV} \\ \hline 
 $\quad\nu\quad$ 
 &  $\quad\xi^{\rm NLL}(\nu)\quad$ 
 & $\quad\xi^{\rm NNLL}_{\rm nm}(\nu)\quad$
 &  $\quad\xi^{\rm NLL}(\nu)\quad$ 
 & $\quad\xi^{\rm NNLL}_{\rm nm}(\nu)\quad$ \\ \hline\hline
 $1.0$ & $0.0000$ & $0.0000$ & $0.0000$ & $0.0000$ \\ \hline
 $0.8$ & $0.0069$ & $0.0041$ & $0.0308$ & $0.0425$ \\ \hline
 $0.6$ & $0.0157$ & $0.0104$ & $0.0712$ & $0.1304$ \\ \hline
 $0.4$ & $0.0274$ & $0.0216$ & $0.1335$ & $0.4537$ \\ \hline
 $0.2$ & $0.0435$ & $0.0512$ &  &  \\ \hline
\end{tabular}
\end{center}
{\caption{
Numerical values for  $\xi^{\rm NLL}(\nu)$ and
$\xi^{\rm NNLL}_{\rm nm}(\nu)$. The values for $m$ are pole masses. 
The numbers are obtained by evaluation of the analytic results using 
four-loop running for $\alpha_s$ and taking 
$\alpha_s^{(5)}(175\,\mbox{GeV})=0.107$ and 
$\alpha_s^{(4)}(4.8\,\mbox{GeV})=0.216$ 
as input.
}
\label{tabcompare} }
\end{table}
In Tab.\,\ref{tabcompare} the values for $\xi^{\rm NLL}(\nu)$ and
$\xi^{\rm NNLL}_{\rm nm}(\nu)$ are displayed for different $\nu$ for the top
and the bottom quarks. For top quarks we find that the NNLL non-mixing
contributions are of the same size as the NLL terms for the relevant region
$\nu\sim v\simeq 0.2$. Here, the new NNLL order corrections shift $c_1$ by about
$+5\%$, which is substantial considering that the total cross section contains
$c_1^2$. The shift is dominated by the contributions coming from the ultrasoft
diagrams shown in Fig.\,\ref{figusoftcorrelators}.
However, it is not yet possible to draw definite phenomenological conclusions
for the normalization of the top threshold cross section in $e^+e^-$
collisions from this result, because the yet unknown NNLL mixing 
corrections could be sizeable as well. It is therefore an important future
task to determine the mixing contributions to the NNLL anomalous dimension of
$c_1$.  

For bottom quarks the NNLL non-mixing contributions are several
times larger than the NLL terms for the relevant region $\nu\sim v\approx
0.3$--$0.4$. This is not unexpected because for bottomonium systems the
binding energy $\sim m v^2$ is already of order $\Lambda_{\rm QCD}$. Thus our
result seems to affirm that for $b\bar b$ states non-perturbative effects have
a rather strong influence, and that the vNRQCD description ceases to work even
for the ground state.\,\cite{Hoang4}

\section{Determination of the Total Cross Section}

The total cross section for $e^+e^-\to \gamma^*, Z^*\to t\bar t$ at threshold
at NNLL order has the form 
\begin{eqnarray}
  \sigma_{\rm tot}^{\gamma,Z}(s) = \frac{4\pi\alpha^2}{3 s} 
  \Big[\, F^v(s)\,R^v(s) +  F^a(s) R^a(s) \Big] \,,
\label{totalcross}
\end{eqnarray}
where $F^{v,a}$ are trivial functions of the electric charges and the isospin
of the electron and the top quark and of the weak mixing angle. At NNLL order
the vector and axial-vector $R$-ratios have the form 
\begin{eqnarray} \label{Rveft}
 R^v(s) & = & \frac{4\pi}{s}\,
 \mbox{Im}\Big[\,
 c_1^2(\nu)\,{\cal A}_1(v,m,\nu) + 
 2\,c_1(\nu)\,c_2(\nu)\,{\cal A}_2(v,m_t,\nu) \,\Big] \,,
\\[4mm] \label{Raeft}
 R^a(s) & = &  \frac{4\pi}{s}\,
 \mbox{Im}\Big[\,c_3^2(\nu)\,{\cal A}_3(v,m_t,\nu)\,\Big] \,,
\end{eqnarray}
where the time-ordered products of the effective theory currents read
($\hat{q}\equiv(\sqrt{s}-2m_t,0)$)
\begin{eqnarray}
 {\cal A}_1 &=& i\,
 \sum\limits_{\mbox{\scriptsize\boldmath $p$},\mbox{\scriptsize\boldmath $p'$}}
 \int\! d^4x\: e^{i \hat{q} \cdot x}\:
 \Big\langle \,0\,\Big|\, T\, \O{p}{1}(x){\Od{p'}{1}}(0)\,
 \Big|\,0\,\Big\rangle \,, \nn
\\[2mm]
 {\cal A}_2 &=&
 \frac{i}{2}\, 
 \sum\limits_{\mbox{\scriptsize\boldmath $p$},\mbox{\scriptsize\boldmath $p'$}}
 \int\! d^4x\: e^{i \hat{q}\cdot x}\:
 \Big\langle \,0\,\Big|\,
 T\, \Big[ \O{p}{1}(x){\Od{p'}{2}}(0)+\O{p}{2}(x){\Od{p'}{1}}(0)\,
 \Big] \Big|\,0\,\Big\rangle \,, \nn
\\[2mm]
 {\cal A}_3 &=& i\, 
 \sum\limits_{\mbox{\scriptsize\boldmath $p$},\mbox{\scriptsize\boldmath $p'$}}
 \int\! d^4x\: e^{i \hat{q}\cdot x}\:
 \Big\langle \,0\,\Big| \,T\, \O{p}{3}(x){\Od{p'}{3}}(0)\Big|\,0\, \Big\rangle \,.
\end{eqnarray} 
The correlators ${\cal A}_i$ can be written in a compact form in terms of
non-relativistic  zero-distance Greens functions. For the convention for the
$1/(m|\bmk|)$-type potentials employed in Ref.\,\cite{Hoang6} they have the form
($v=((\sqrt{s}-2m_t+i\Gamma_t)/m_t)^{1/2}$),  
\begin{eqnarray}
\lefteqn{
{\cal A}_1(v,m,\nu)
 \, = \,
 6 \,N_c\,\Big[\,
 G^c(v,m,\nu)
 + \Big({\cal{V}}_2(\nu)+2{\cal{V}}_s(\nu)\Big)\, 
 \delta G^\delta(v,m,\nu)}
\nonumber \\[2mm] & &
+ \:{\cal{V}}_r(\nu)\,\delta G^r(v,m,\nu) 
+ \delta G^{\rm kin}(v,m,\nu) 
\nn \\[2mm] & &
- \:C_AC_F\,\alpha_s^2(m\nu)\,\delta G^k_{\rm CACF}(v,m,\nu) 
+\frac{C_F^2}{2}\,\alpha_s^2(m\nu)\,\delta G^k_{\rm CF2}(v,m,\nu)\qquad
\nn \\[2mm] & &
+ \:\alpha_s^2(m\nu){\cal V}_{k1}(\nu)\,\delta G^{k1}(v,m,\nu) 
+   \alpha_s^2(m\nu){\cal V}_{k2}(\nu)\,\delta G^{k2}(v,m,\nu)
\,\Big]\,,
\nonumber\\[2mm] 
\lefteqn{
{\cal A}_2(v,m,\nu) \, = \, {v^2}\,{\cal A}_1(v,m,\nu) \,,
\quad
{\cal A}_3(v,m,\nu) \, = \, \frac{4 \,N_c}{m_t^2}\, G^1(a,v,m,\nu)
\,.}
\label{NNLLcrosssection}
\end{eqnarray}
The Greens function $G^c$ contains the sum of $1/\bmk^2$ potentials from
Eqs.\,(\ref{VCoulombspoft}) and (\ref{vNRQCDpotential}) and was computed
numerically in Ref.\,\cite{Hoang4,hmst}. (See also Ref.\,\cite{Jezabek}.) 
All other Greens functions were obtained 
analytically in dimensional regularization. The terms $\delta G^{\delta,r}$
arise from a single insertions of the $1/m_t^2$ potentials in
Eq.\,(\ref{vNRQCDpotential}) and  $\delta G^{\rm kin}$ from an insertion of
the kinetic energy $\bmp^4/(8 m_t^3)$.~\cite{Hoang4}  The terms $\delta G^{\rm
  k,k1,k2}$ come from a single 
insertion of the $1/(m|\bmk|)$-type potentials and their expressions were obtained
in Refs.\,\cite{Hoang5,Hoang6}. The P-wave Greens function $G^1$ was also
obtained in 
Ref.\,\cite{Hoang4}. The Greens functions contain UV subdivergences which are
removed by the renormalization constant of the current $\O{p}{1}$. They still
contain overall divergences which, however, are not contained in their
absorptive part for stable quarks. 

I have displayed the result in the pole mass scheme.   
Since the pole mass is plagued by a renormalon ambiguity of order
$\Lambda_{\rm QCD}$, which causes an instability in the prediction of the
cross section, it is mandatory to switch to a ``threshold mass'' mass
definition such as the kinetic, the PS of the 1S mass for
phenomenological examinations.~\cite{Hoang3} I will use the 1S mass
definition~\cite{Hoang2,Hoang7} for the discussions in the next section.  
The Greens functions also depend on the renormalization parameter $\nu$. For
$\nu$ of order $|v|$, the top quark velocity, the Greens function do not contain
any large logarithmic terms and all logarithms are summed into the Wilson
coefficients of the potentials and the currents by the renormalization group
equations discussed in the previous section. Typically, one chooses $\nu\simeq
0.15-0.2$, which corresponds to a momentum scale $m_t\nu\simeq 25-35$\,GeV and
an energy scale $m_t\nu^2\simeq 4-7$\,GeV.     

Recently, in Ref.\,\cite{Hoang6} the order $\alpha_s^3\ln\alpha_s$
corrections to the heavy quarkonium partial width into a lepton pair were
computed  in the fixed-order expansion using an asymptotic  
expansion of QCD diagrams close to threshold\,\,\cite{Kniehl2}. (For 
a discussion of the behavior of the perturbative series in the fixed-order
expansion I refer to Ref.\,\cite{Kniehl2}.) 
A summation of logarithms is not contained in that work, but the result can be 
used as a non-trivial cross check for the summations contained in the NNLL
vector R-ratio $R^v$ of Eq.\,(\ref{Rveft}). Expanding out the summations
contained in the Wilson coefficients for energies on the bound state poles one
can determine the analogous order $\alpha_s^3\ln\alpha_s$ terms.
The result agrees with the updated result of Ref.\,\cite{Kniehl2}.

\section{Discussion}

Let me now turn to how the NNLL non-mixing contributions 
affect the vector-current-induced
top threshold cross section $R^v$ numerically. 
\begin{figure}[t] 
\begin{center}
\leavevmode
\epsfxsize=3.8cm
\leavevmode
\epsffile[230 585 428 710]{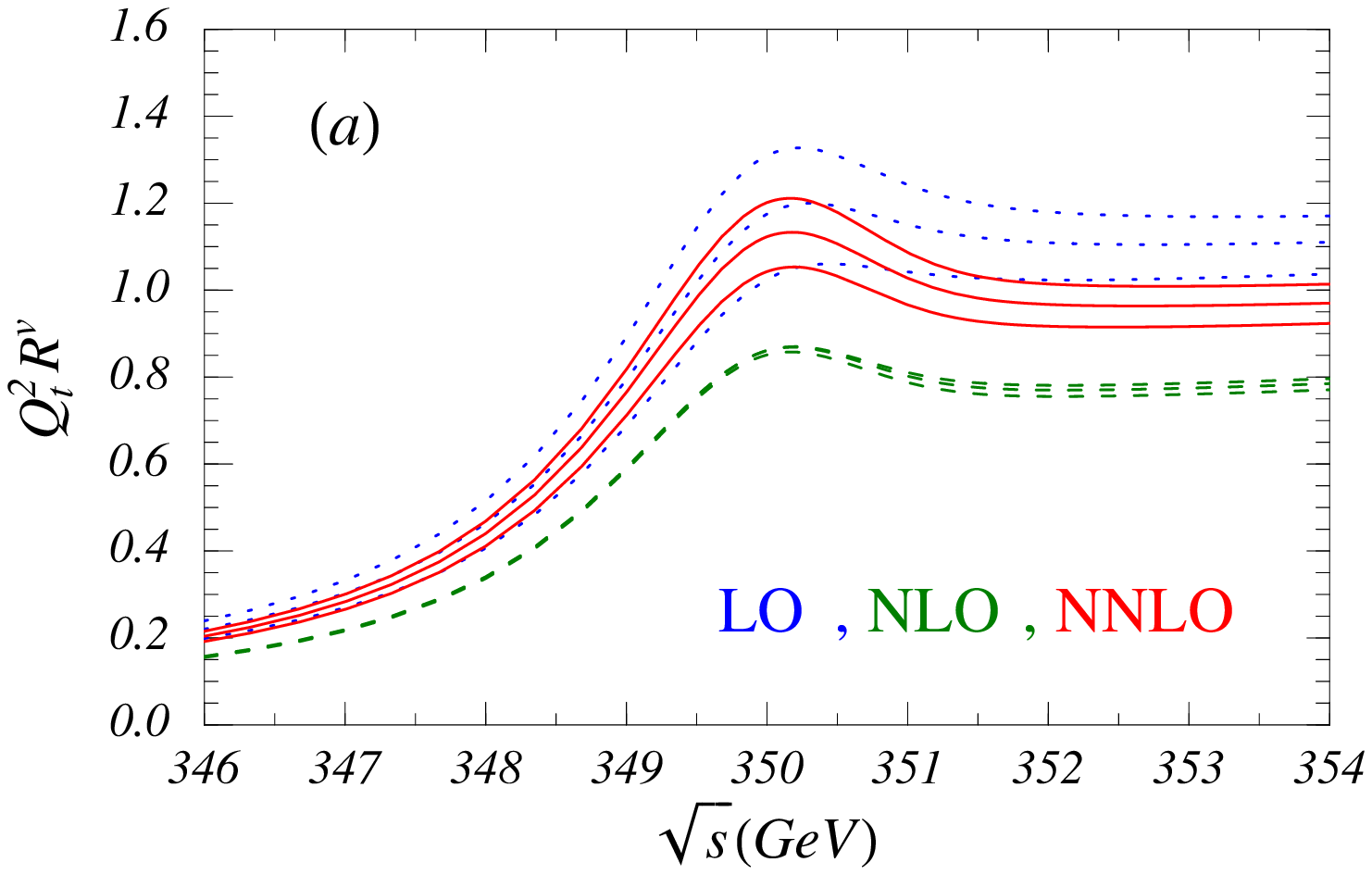}
\\[2.9cm]
\leavevmode
\epsfxsize=3.8cm
\leavevmode
\epsffile[230 585 428 710]{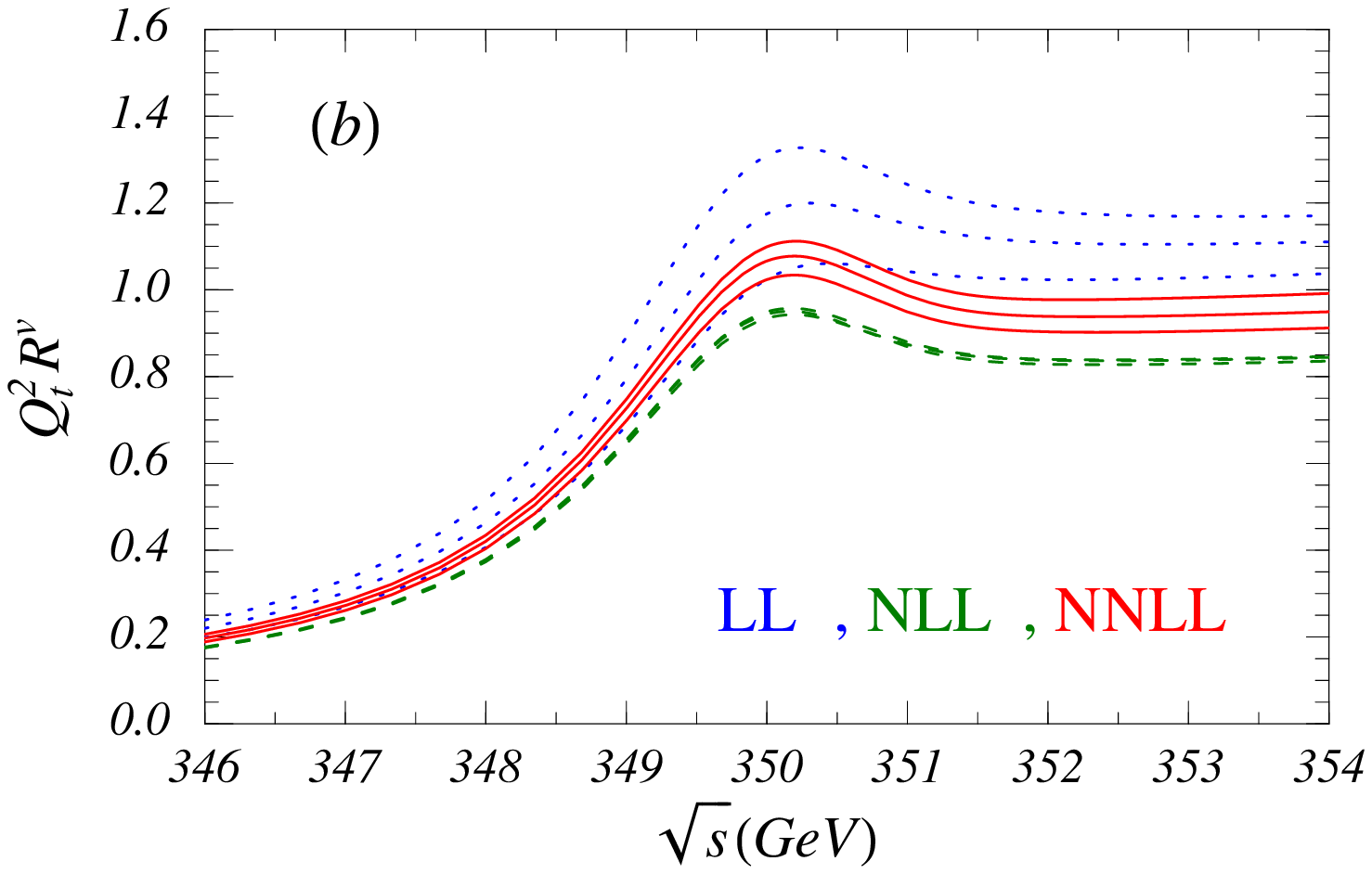}
\vskip  2.6cm
 \caption{ Panel a) shows the results for $Q_t^2 R^v$ with $M^{\rm
 1S}=175$\,GeV and $\Gamma_t=1.43$\,GeV in fixed-order perturbation theory at
 LO (dotted lines), NLO (dashed lines) and NNLO (solid lines). 
 Panel b) shows the results for $Q_t^2 R^v$ with the same parameters 
 in renormalization group improved perturbation theory at
 LL (dotted lines), NLL (dashed lines) and NNLL (solid lines) order. 
 For each order curves are plotted for $\nu=0.15$, $0.20$, and $0.3$. 
 The effects of initial state radiation, beamstrahlung and the beam energy
 spread at a $e^+e^-$ collider are not included.
\label{figtopplots} 
}
 \end{center}
\end{figure}
In Fig.\,\ref{figtopplots}b I have displayed $Q_t^2 R^v$ up to NNLL order.  The
curves show the LL (dotted blue lines), NLL (dashed green lines) and NNLL (solid
red lines) cross section for $\nu=0.15, 0.2$ and $0.3$.  Figure\,\ref{figtopplots}a
shows the corresponding results in fixed-order perturbation theory.
Compared to earlier analyses
where the NNLL non-mixing contribution were not yet
accounted for~\cite{Hoang4,hmst,Hoang5}, 
the NNLL cross section is shifted upwards by about $+10\%$. On the other hand, the
scale variation of the NNLL result is moderate and about $\pm 3\%$ 
for the variation of $\nu$ I have used in this analysis.   
Compared to the fixed-order results with the same scales the improvement is
substantial, particularly around the peak position and for smaller
energies, but not as dramatic as concluded from our earlier analyses 
when the NNLL non-mixing contributions were not yet included. Although it
appears premature to me to draw any definite 
conclusions from this result for the determination of the top Yukawa coupling,
the strong coupling or the top quark width, because the mixing contributions
are still not computed, it is prudent to say that presently the error estimate
of $\pm 3\,\%$  made in Refs.\,\cite{Hoang4,hmst,Hoang5} cannot be upheld 
and should
be enlarged to $\delta\sigma_{t\bar t}/\sigma_{t\bar t}\simeq \pm 6\,\%$ due
to the relatively large shift between the NLL and the NNLL order results.
As far as the determination of the top quark mass from a threshold scan is
concerned, the new NNLL order results are not expected to affect the prospects
of a determination with $\delta m_t\sim 100$~MeV since already for the
fixed-order NNLO and the earlier NNLL order results the conclusions for the
top mass 
determination from simulation studies were quite
similar.\,\cite{TTbarsim,Peralta1}  

\section{Conclusion}

In this talk I have discussed the ingredients needed to carry out a
renormalization group improved QCD computation of the total cross section
$\sigma(e^+e^-\to t\bar t)$ in the threshold regime. In renormalization group
improved perturbation theory QCD logarithms of the top quark velocity are summed
up to all order in $\alpha_s$ according to the schematic expansion shown in
Eq.\,(\ref{RNNLLorders}). The summations can be obtained within an effective
theory where all degrees of freedom that can fluctuate close to their 
mass-shell are represented as fields and all off-shell fluctuations are
integrated out. All logarithmic terms are associated to UV divergences in the
effective theory and the summation of logarithms is achieved by solving the
renormalization group equations of the couplings and coefficients of the
effective theory.  At present all ingredients for a full NNLL order
prediction of the total threshold cross section are known except for the 
NNLL order result of $c_1$, the coefficient of the dominant current that
produces the top pair in $e^+e^-$ annihilation, which is only fully known at
NLL order.
I have presented new results for the NNLL non-mixing contributions of
anomalous dimension of the coefficient $c_1$. Numerically,
the NNLL non-mixing contributions to $c_1$ are of the same size as the NLL
contributions and shift the total cross section by about $+10\%$. 
Including the new results the present theoretical uncertainty for
the normalization of the total cross section is about $\pm 6\%$. The new result
does not affect the prospects for a precise determination of the top quark
mass from a threshold scan based on earlier work, but it does affect
the prospects for the determination of the top Yukawa coupling, the strong
coupling and the top quark width. However, it is premature to draw
definite conclusion as long as the NNLL mixing contributions have not yet been
determined. 

\section{Acknowledgments}
I would like to thank A.~Manohar, I.~Stewart and T.~Teubner for their
collaboration on the results presented here and T.~Teubner for comments to the
manuscript.


\end{document}